\documentclass{article}
\usepackage{graphicx} 
\usepackage{physics,mathtools,amsmath,amssymb,amsfonts,latexsym,amscd,euscript,url,dsfont}
\usepackage{tikz}
\usepackage[a4paper,left=2.5cm,right=2.5cm,top=2.5cm,bottom=2.5cm]{geometry}
\usepackage{mathrsfs}
\usepackage[normalem]{ulem}
\usepackage{cancel}
\usepackage{wrapfig}
\usepackage[affil-it]{authblk}
\newtheorem{theorem}{Theorem}

\usepackage{circuitikz}
\usepackage[sort&compress,numbers]{natbib}
\usepackage{appendix}
\usepackage{comment}

\usepackage[dvipsnames]{xcolor}
\usepackage{ulem}
\usepackage{jheppub}

\begin{document}
\title{{Teleparallel gravity from the principal bundle viewpoint}}
\author[1]{Sebastian Brezina,}
\author[2]{Eugenia Boffo,}
\author[1]{Martin Kr\v{s}\v{s}\'ak}
\affiliation[1]{Department of Theoretical Physics, Faculty of Mathematics,
Physics and Informatics, Comenius University in Bratislava, 84248, Slovak Republic}
\affiliation[2]{Mathematical Institute, Faculty of Mathematics and Physics, Charles University, Prague 186 75, Czech Republic}
\emailAdd{sebastian.brezina@fmph.uniba.sk}
\emailAdd{boffo@karlin.mff.cuni.cz}
\emailAdd{martin.krssak@fmph.uniba.sk}
\date{\today}
\abstract{We examine whether the Teleparallel Equivalent of General Relativity (TEGR) can be formulated as a gauge theory in the language of connections on principal bundles.
    We argue in favor of using either the affine bundle with the Poincaré group or, equivalently, the orthonormal frame bundle with the Lorentz group as the structure group. Following the framework of Trautman--where gauge symmetries are determined using the absolute elements--we set to identify the absolute elements and gauge symmetries of TEGR.  
    The triviality of the field equations for metric teleparallel connection raises the question of whether it should be treated as a dynamical variable. If the connection is treated as dynamical, then the only absolute element is the canonical 1-form of the frame bundle, and the gauge group of TEGR is the full diffeomorphism group. If the connection is considered as non-dynamical, we show that its treatment as an absolute element leads to problems of not being able to determine the gauge group and its possible non-uniqueness. On the other hand, if the connection is treated as a non-dynamical variable but not as an absolute element, we again recover the whole diffeomorphism group as the gauge group of TEGR.  
    }
\maketitle

\section{Introduction}  
The gauge theory paradigm is a powerful unifying framework for describing the fundamental interactions of nature \cite{Quigg:2013ufa,Eguchi:1980jx}. The notion of gauge theory refers to a theory with a Lagrangian that possesses some local symmetry and is often viewed to represent some redundant degrees of freedom. The primary example is the case of electromagnetism, exhibiting a local $U(1)$ symmetry, and the strong and weak nuclear forces described by the non-Abelian gauge theories of the Yang-Mills type.
A major step in our understanding of gauge theories was the realization, primarily by Trautman \cite{Trautman:1970cy}, as well as others \cite{Lubkin:1963zz,Hermann}, that gauge theories of the Yang-Mills type can be understood using the geometry of principal fiber bundles, developed entirely independently as a field of mathematics \cite{KN}. This has provided a new systematic way of understanding classical gauge theories while putting them on solid mathematical foundations.

An interesting question is whether the remaining fundamental interaction--gravity--can be described as a gauge theory as well. 
The standard theory of gravity is general relativity, which is built on the equivalence principle stating that the laws of special relativity hold only locally, and the principle of general covariance, from which it follows that not all components of the metric are uniquely given. The contracted Bianchi identity for the Riemann curvature tensor--related to the second Noether theorem and covariant conservation laws \cite{Noether:1918zz}--reduces the number of degrees of freedom of the metric, which is also supported by the constraint analysis in the Hamiltonian formulation \cite{Arnowitt:1959ah,Wald:1984rg}. This resembles the general ideas of a gauge theory: some global physical laws are promoted to hold only locally, and we have redundant degrees of freedom in the description of a gravitational field. So, can general relativity, or more generally, a theory of gravity, be consistently formulated as a gauge theory, and, if yes, what is its gauge group?

The gauge approach to gravity goes back almost 70 years to the pioneering works of Utiyama, who proposed to view gravity as a gauge theory of the Lorentz group \cite{Utiyama:1956sy}, which was further extended by Kibble \cite{Kibble:1961ba} to include translations and view gravity as a gauge theory of the whole Poincaré group \cite{Hehl:1976kj,Hehl:1994ue,Blagojevic:2002du,Obukhov}.

On the other hand, a rather different approach was taken by Hayashi and Nakano \cite{Hayashi:1967se}, and Cho \cite{Cho:1975dh}, who have argued that general relativity can be viewed as a gauge theory of translations only since the Einstein-Hilbert action can be expressed--up to boundary terms--using the translational field strength \cite{Cho:1975dh}. It was then shown by  Hayashi \cite{Hayashi:1977jd} that this is equivalent to the teleparallelism model originally proposed by Einstein in the late 1920s as an attempt for unified field theory \cite{Einstein1929a}, which was later revisited by Møller \cite{Moller1961}\footnote{For a review of different approaches to teleparallel gravity with some historical background,  see \cite{Krssak:2024xeh}.}.  This case, when the action is dynamically equivalent to general relativity but expressed in terms of the metric teleparallel geometry, is now usually referred to as the Teleparallel Equivalent of General Relativity (TEGR) \cite{Aldrovandi:2013wha,Maluf:2013gaa,Krssak:2018ywd}, and is the starting point for various extended models of gravity popular in recent years \cite{Ferraro:2006jd,Bahamonde:2021gfp}. 

The main question here is to what extent the modern version of TEGR can be understood as a gauge theory, which remains a subject of debate. On one side, some argue that it is a gauge theory of translations in Cho's sense \cite{Aldrovandi:2013wha,Krssak:2018ywd,Pereira:2019woq}, while others question this interpretation, as it relies on a series of identifications that may be considered problematic. These include the identification of translations with general coordinate transformations, translational gauge potentials with frame fields, and translational field strength with torsion. This critique was recently raised in a series of papers \cite{Fontanini:2018krt,LeDelliou:2019esi,Huguet:2021roy}, but similar concerns were already voiced in the 1980s by Dass \cite{Dass:1984qk}, and are closely related to issues raised by Sardanashvily in the context of Poincaré gauge theories \cite{Sardanashvily:1983xbn,Sardanashvily:1984dx,Ivanenko:1987ka,Sardanashvily:2002vq}.

We explore the gauge structure of TEGR from the principal bundle perspective following the approach of Trautman \cite{Trautman:1979fg,Trautman-G}, where we start with a principal bundle with structure group $G$, and the gauge groups are determined using the absolute elements of the theory, i.e., geometric objects that do not have field equations. In the case of electromagnetism or gauge theories of the Yang-Mills type over Minkowski space,  one of the absolute elements is the Minkowski metric, and we recover the ''common physicist's" notion of a gauge symmetry as some global symmetry "promoted" to a local one. The advantage of Trautman's framework is that it can be straightforwardly applied to the case of gravity, where the absolute element is the canonical 1-form living on the frame bundle\footnote{We remark that a slightly different, but equivalent, approach to gauge gravitational theories was taken by Sardanashvily \cite{Sardanashvily:1984dx,Ivanenko:1987ka,Sardanashvily:2002vq,Sardanashvily:2011rp}. See \cite{Percacci:1984bq,Percacci:2009ij} and the more recent \cite{Francois:2025jro} as well.}.
 In general relativity and other classical gauge theories of gravity based on a dynamical metric tensor and a dynamical linear connection, the gauge group is the group of diffeomorphisms and the structure group is $GL(4,\mathbb{R})$, which, due to the presence of a metric tensor $g$, spontaneously breaks to $SO(1,3)$.

We apply this framework to the case of TEGR, where we first discuss both the translational and Poincaré groups as the candidates for the structure group. We then encounter the well-known problem of the teleparallel connection not being determined by the field equations \cite{Obukhov:2002tm,Krssak:2015lba,Golovnev:2017dox, Krssak:2018ywd}, which may imply that teleparallel theories do not straightforwardly fit into Trautman's framework for classical gauge theories, where the connection is required to be dynamical-- to have field equations. We first explore the case when the connection is treated as a dynamical variable, which leads to the full diffeomorphism symmetry of TEGR and its interpretation as a classical gauge theory within Trautman's framework. We then also explore the possibility when the connection is considered to be non-dynamical, which leads us to the question of whether the metric teleparallel connection should be treated as an absolute element or not. We argue that in the former case, this results into a problem where the gauge group of the theory becomes some subgroup of spacetime diffeomorphisms, which is not explicitly determined. On the other hand, if the connection is allowed to be non-dynamical but not treated as an absolute element, we recover again that the gauge group is the full group of spacetime diffeomorphisms, and hence the theory is equivalent to general relativity.

The paper is organized as follows: 
In Section~\ref{P1}, we summarize Trautman's approach to classical gauge theories and demonstrate it on the examples of Yang-Mills theories and gravitational theories with dynamical connection. In Section~\ref{P2}, we apply this approach to TEGR and discuss both translational and Poincaré groups as the structure group candidates. The main result is presented in Section~\ref{p2-s22}, where we argue in favor of the Poincaré group and study the gauge interpretation of TEGR from the viewpoint of Trautman's framework and its possible modification. In Section~\ref{subsec-bridge}, our results are presented from the viewpoint of formalism used in Ref.~\cite{Aldrovandi:2013wha,Krssak:2018ywd,Pereira:2019woq} to better explain the relation to our approach. 
We have included a brief summary of the mathematical formalism used throughout the paper in  Appendices, where we have summarized the most important concepts and terminology from the theory of connections following primarily Fecko \cite{Fecko:2006zy} {(chapters 17,19-21)}, with additional details available in \cite{KN} and \cite{Trautman:1985wm}. 

%%%%
%%%%
%%%%
\section{Classical gauge theories}\label{P1}
We start by recalling Trautman's definition of classical 
gauge theories \cite{Trautman:1981vr}: \emph{A classical gauge theory is any physical theory which includes among its dynamical variables a connection on principal $G$-bundle over spacetime}. 
To construct a gauge theory, we have to necessarily specify \cite{Trautman:1981fd}: 
\begin{enumerate}
    \item a structure group $G$; 
    \item the type of particles which couple to gauge fields (this is done by choosing functions $\Phi$ of type $\rho$);
    \item the form of the field equations (usually following from a variational principle). 
\end{enumerate}

In classical gauge theories, we work with the following objects: a \emph{gauge field} $\mathcal{A}$, the \emph{gauge field strength} $\mathcal{F}$, and a \emph{matter field} $\phi$. In theory of connections on principal $G$-bundles \cite{KN} (see Appendix \ref{AppA}) we work with objects: a \emph{connection $1$-form} $\omega$, the \emph{curvature $2$-form} $\Omega$, and a \emph{function $\Phi$ of type $\rho$}. It turns out \cite{Fecko:2006zy} that all these objects are related as
\begin{equation}\label{P1-1}
    \mathcal{A}=\rho'\left(\sigma^*\omega\right), \quad\quad \mathcal{F}=\rho'\left(\sigma^*\Omega\right), \quad\quad \phi=\sigma^*\Phi,
\end{equation}
where $\rho'$ is the derived representation of the representation $\rho$ of $G$ and $\sigma$ denotes a local section of the principal $G$-bundle.

The definitions of gauge groups were provided in \cite{Trautman:1979fg,Trautman-G,Trautman:1981qm,Mayer:1981qk}. We follow here primarily \cite{Trautman-G}.
Let us have a principal $G$-bundle $\pi: P\rightarrow M$. A diffeomorphism\footnote{A diffeomorphism is a smooth bijective map between manifolds $M$ and $N$. In the case $N=M$, diffeomorphisms constitute an $\infty$-dim. group denoted as $\text{Diff}~M$.} 
$u: P\rightarrow P$ is an \emph{automorphism} of the principal $G$-bundle  if there is a diffeomorphism $v: M\rightarrow M$ such that $\pi\circ u=v\circ \pi$ and $u$ satisfies $u\circ R_g=R_g\circ u$ for all $g\in G$. The set $\text{Aut}~P$ of all automorphisms of $P$ is an $\infty$-dim. group under the composition of maps. The automorphism $u$ uniquely determines the diffeomorphism $v$, and there is a homomorphism $j$ between $\text{Aut}~P$ and $\text{Diff}~M$ groups. An automorphism $u$ is called \emph{vertical} if $v=id_M$. The set of vertical automorphisms $\text{Aut}_0~P$ is a normal subgroup of $\text{Aut}~P$. 

\emph{Absolute elements (non-dynamical elements)} of a classical gauge theory are those geometric objects in the gauge theory that do not have field equations \cite{TrautmanGR}, i.e.~they are external and exist independently of it. By definition, the \emph{gauge group} $\mathscr{G}$ of a classical gauge theory containing absolute elements consists of all automorphisms of $P$ which preserve them, thus $\mathscr{G}\subset\text{Aut}~P$. The elements of $\mathscr{G}$ are called \emph{gauge transformations}. The \emph{pure gauge group} $\mathscr{G}_0$ is defined as $\mathscr{G}_0:=\mathscr{G}\cap \text{Aut}_0~P$, i.e. it consists of gauge transformations which are \emph{vertical}. The elements of $\mathscr{G}_0$ are called \emph{pure gauge transformations}.

Gauge and pure gauge transformations act on local sections of $P$. Let $\sigma:\mathcal{U}\rightarrow P$ be a local section of $P$, where $\mathcal{U}\subset M$ is an open subset. Then a gauge transformation gives another local section $\hat{\sigma}:=u\circ\sigma\circ v^{-1}: v(\mathcal{U})\rightarrow P$. It turns out \cite{Mayer:1981qk} that if $u\in \mathscr{G}_0$, then the relation between local sections is fully described by maps $S: \mathcal{U}\rightarrow G$ given as $\hat{\sigma}(x)=R_{S(x)}\sigma(x)$. These maps constitute the group $G^{\mathcal{U}}$ under the composition of maps \cite{Fecko:2006zy}. Constant elements of $G^{\mathcal{U}}$ constitute a finite-dimensional subgroup that is isomorphic to the structure group $G$. Let us define \cite{Fecko:2006zy} the matrix representation of $S$ as $B:=\rho\circ S: \mathcal{U}\rightarrow GL(N,\mathbb{R})$, where $N=\text{dim}(V,\rho)$, i.e. $B(x):=\rho(S(x))$, then pure gauge transformations yield known transformation laws for fields $\mathcal{A}, \mathcal{F}$ and $\mathcal{\phi}$ (for compact notation we write just $B\equiv B(x)$)
\begin{equation}
    \hat{\mathcal{A}}=B^{-1}\mathcal{A}B+B^{-1}dB, \quad\quad \hat{\mathcal{F}}=B^{-1}\mathcal{F}B, \quad\quad \hat{\phi}=B^{-1}\phi. 
\end{equation}

Let us now demonstrate this rather abstract framework on the examples of Yang-Mills theory, electromagnetism, and gravitational theories with dynamical connections {following \cite{Trautman-G}}. This helps us clarify differences in the terminology used in physics, which causes misunderstandings in the case of gravitational theories.

\emph{Yang-Mills theory over Minkowski spacetime} is a classical gauge theory built using a principal $G$-bundle over Minkowski space $\left(\mathbb{R}^4,\eta,\nabla_{LC}\right)$, where the structure group $G$ is a  compact and semi-simple Lie group. {In the absence of sources, $\mathbb{R}^4$ is the base manifold, which is contractible and hence the principal bundle is trivial}, i.e.~$P=\mathbb{R}^4\times G$, see Theorem \ref{T1} in  Appendix \ref{AppA}. 
The field equations of Yang-Mills theory are derived by varying the action
\begin{equation}\label{YMA}
    \mathcal{S}_{YM}[\mathcal{A}]:=-\frac{1}{2}\int \text{Tr}\left\{\mathcal{F}\wedge*\mathcal{F}\right\},
\end{equation} 
where $*$ is the Hodge star operator.

We take the Minkowski metric $\eta$ to be the absolute element since $\eta$ does not have field equations, but rather is a consequence of the principles of special relativity \cite{Trautman-G}. The gauge group is then determined as a group of automorphisms of the principal $G$-bundle $P$ preserving  $\eta$
\begin{equation}
\mathscr{G}=\{u\in\text{Aut}~P;~v^*\eta=\eta,~ v=j(u)\},
\end{equation}
which can be shown to be given by $\mathscr{G}=\mathscr{G}_0\rtimes\mathcal{P}$ , where $\mathcal{P}=SO(1,3)\ltimes \mathbb{T}(4)$ is the \emph{Poincaré group}.  The pure gauge group is   $\mathscr{G}_0=\left\{S:\mathbb{R}^4\rightarrow G\right\}$,  i.e., to each point of Minkowski spacetime is associated some element of $G$.

Similarly, in the case of \emph{Maxwell's electromagnetism}, which  is a classical gauge theory of the Yang-Mills type over Minkowski space with the Abelian structure group $G=U(1)$ and the action
\begin{equation}\label{EMA}
	\mathcal{S}_{EM}[A]:=-\frac{1}{2}\int \mathcal{F}\wedge*\mathcal{F},
\end{equation}
the absolute element is once again the Minkowski metric tensor $\eta$, and we obtain the gauge group to be $\mathscr{G}=\mathscr{G}_0\rtimes \mathcal{P}$, where $\mathscr{G}_0=\left\{S:\mathbb{R}^4\rightarrow U(1); ~ S(x)=e^{i\alpha(x)}\right\}$, while $G=\left\{S:\mathbb{R}^4\rightarrow U(1); ~ S(x)=e^{i\alpha}\right\}$.

We can observe that the pure gauge group corresponds to what is commonly referred to as the group of \textit{local gauge transformations} in the physics literature, while the elements of the structure group $G$ can be understood as the \textit{global gauge transformations}. Therefore, in the case of classical gauge theories of Yang-Mills type over Minkowski spacetime, we recover the common notion that local gauge transformations are obtained by ``localizing" the global gauge transformations.  Since the knowledge of $G$ is sufficient in this special case, this explains why the structure group $G$ is often referred to simply as the gauge group.

The important insight of Trautman's framework is that the gauge group $\mathscr{G}$ is determined not only from the structure group $G$, but also from the knowledge of the manifold $M$ and the absolute elements.  While in the special case of theories of  Yang-Mills type over Minkowski spacetime, this corresponds to the common picture of ``localizing" the global gauge transformations, it is more general and allows us to treat other situations as well, including theories of Yang-Mills type with the non-trivial principal bundles and--most importantly--gravity.

An example of the former is electromagnetism with magnetic monopoles, which can be described as a gauge theory with $G=U(1)$ over the spacetime manifold homeomorphic to $\mathbb{R}^2\times S^2$ rather than $\mathbb{R}^4$, and the underlying principal $U(1)$-bundle assumed to be non-trivial \cite{Trautman:1985wm}. The interesting insight here is that when comparing the $U(1)$-bundles $\pi: \mathbb{R}^4\times U(1) \rightarrow \mathbb{R}^4$ and $\pi': P'\rightarrow\mathbb{R}^2 \times S^2$, we can notice that their respective automorphisms groups are different, and hence the gauge group depends on the specific $(P,M,\pi)$ under inspection.

\emph{Gravitational theories} are understood as dynamical theories of a metric tensor $g$, with a dynamical linear connection $\nabla$ over a 4-dimensional orientable Lorentzian manifold $M$,  i.e. described geometrically as $(M,g,\nabla)$.  So we work with the frame bundle $LM$ with a $gl(4,\mathbb{R})$-valued linear connection $\omega$. On $LM$ (or\footnote{The existence of $g$ with the signature $(1,3)$ spontaneously breaks the $GL(4,\mathbb{R})$ symmetry to $O(1,3)$, and gives rise to the orthonormal bundle $OM$ which is a principal $O(1,3)$-bundle. If we want a linear connection $\omega$ to be defined only on $OM$, then it is sufficient and necessary to require the non-metricity tensor $\nabla g$ to vanish, or equivalently $\omega$ must be just $o(1,3)$-valued, for more details, see chapter 20.5 of \cite{Fecko:2006zy}.} 
on $OM$) we additionally have the \emph{canonical 1-form $\theta$}, see \eqref{LM1}, which becomes the absolute element of gravitational theories
since it does not have field equations and is given by the choice of $M$. Therefore, the gauge group is given as 
\begin{equation}\label{Gaugegroup}
    \mathscr{G}=\{u\in\text{Aut}~LM;~u^*\theta=\theta\}
\end{equation}
It turns out \cite{Trautman-G} that $\mathscr{G}$ is isomorphic to $\text{Diff}~M$, and the pure gauge group reduces to the trivial group, i.e. $\mathscr{G}_0=\{id_{LM}\}$. 
Essentially this happens because the automorphisms $u$ of $LM$ preserving $\theta$ are captured by $u=L(v)$, where $v$ is a diffeomorphism of $M$ mapped by the \emph{lift} $L$,
satisfying $j\circ L=id_{\text{Diff}~M} $, 
into an automorphism of $LM$. We will expand upon this argument in subsec. \ref{p2-s22} when addressing the teleparallel equivalent to general relativity as a classical gauge theory with Poincar\'{e} structure group.

This explains the common misunderstanding occurring when the structure group $GL(4,\mathbb{R})$ acting on $LM$ (or\footnote{We are usually interested in parity-preserving theories and hence we consider only $SO(1,3)\subset O(1,3)$.} $SO(1,3)$ on $OM$) is referred to as the gauge group. The reason why we call this a misunderstanding is that it would suggest that local gauge transformations are local $GL(4,\mathbb{R})$ (or local $SO(1,3)$) transformations, but this is not the case here, since the gauge group \eqref{Gaugegroup} does not contain any non-trivial vertical automorphism\footnote{This follows from the definition \eqref{LM1} and the definition of a vertical automorphism. The action of the group of vertical automorphisms on local sections in a region $\mathcal{U}\subset M$ is fully given through the local $GL(4,\mathbb{R})$ group, i.e., it is given as the group of maps $S:\mathcal{U}\rightarrow GL(4,\mathbb{R})$. Thus, no non-trivial local $GL(4,\mathbb{R})$ transformation is part of $\mathscr{G}$.}.

In \emph{general relativity}, whose geometrical structure is $\left(M,g,\nabla_{LC}\right)$, gravity is described as a manifestation of the curvature of the Levi-Civita connection, and the matter-free action reads 
\begin{equation}\label{GRA}
   \mathcal{S}_{GR}[e]:= -\frac{1}{16\pi}\int \hat{\Omega}_{ab}\wedge*\left(e^a\wedge e^b\right),  
\end{equation}
where the notation used is explained in Appendix \ref{AppB}. General relativity can then be understood as a classical gauge theory with the structure group $SO(1,3)$ and the gauge group $\text{Diff}~M$, and no "internal symmetries" are present ($\mathscr{G}_0$ is trivial).

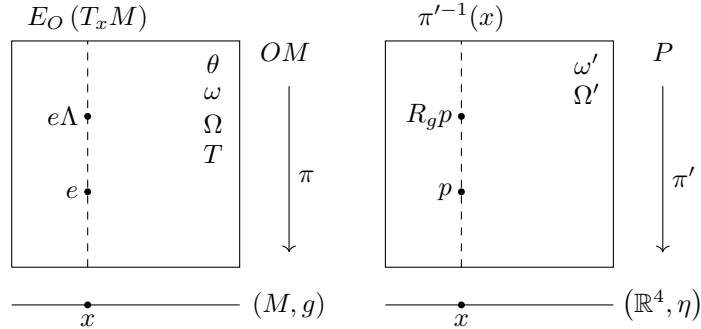
\begin{figure}[!tbp]\label{fig1}
  \centering
  \begin{minipage}[b]{0.3\textwidth}
   \begin{tikzpicture}
\draw[](0,0)rectangle(3,3);
\fill(3.65,-0.5) node[] {$(M,g)$};
\fill(3.6,2.85) node[] {$OM$};
\fill (1,1) circle[radius=1.3pt] node[anchor= east] {$e$};
\fill (1,2) circle[radius=1.3pt] node[anchor= east] {$e\Lambda$};
\fill(1,3) node[anchor= south] {$E_O\left(T_xM\right)$};
\fill(2.65,2.7) node[] {$\theta$};
\fill(2.65,2.3) node[] {$\omega$};
\fill(2.65,1.9) node[] {$\Omega$};
\fill(2.65,1.5) node[] {$T$};
\fill(1,-0.5) circle[radius=1.3pt] node[anchor= north] {$x$};
\fill(3.65,1.2) node[anchor= west] {$\pi$};
\draw[] (0,-0.5) -- (3,-0.5);
\draw[dashed] (1,0) -- (1,3);
\draw[->] (3.65,2.4) -- (3.65,0.2);
\end{tikzpicture}
  \end{minipage}
  %\hfill
  \begin{minipage}[b]{0.25\textwidth}
   \begin{tikzpicture}
\draw[](0,0)rectangle(3,3);
\fill(3.65,-0.5) node[] {$\left(\mathbb{R}^4,\eta\right)$};
\fill(3.65,2.85) node[] {$P$};
\fill (1,1) circle[radius=1.3pt] node[anchor= east] {$p$};
\fill (1,2) circle[radius=1.3pt] node[anchor= east] {$R_gp$};
\fill(1,3) node[anchor= south] {$\pi'^{-1}(x)$};
\fill(1,-0.5) circle[radius=1.3pt] node[anchor= north] {$x$};
\fill(3.65,1.2) node[anchor= west] {$\pi'$};
\draw[] (0,-0.5) -- (3,-0.5);
\draw[dashed] (1,0) -- (1,3);
\draw[->] (3.65,2.4) -- (3.65,0.2);
\fill(2.65,2.7) node[] {$\omega'$};
\fill(2.65,2.3) node[] {$\Omega'$};
\end{tikzpicture} 
  \end{minipage}
  \caption{A geometrical comparison of gauge gravitational theories with vanishing non-metricity tensor (on the left) and gauge theories of the Yang-Mills type over Minkowski space (on the right).}
\end{figure}

Now we go over all the geometrical ingredients again, with the aim to compare the geometrical structures of gravitational theories (with spontaneously broken $GL(4,\mathbb{R})$ symmetry and vanishing $\nabla g$) with gauge theories of the Yang-Mills type over Minkowski space. This should provide a deeper insight into the geometry of classical gauge theories. Figure \ref{fig1} summarizes the crucial geometrical differences between gravitation and gauge theories of the Yang-Mills type over Minkowski space. The former is based on the principal bundle $OM$ \emph{soldered} to the base manifold, while the latter is based on "abstract principal bundles" $P$. Soldering gives rise to the canonical $1$-form $\theta$ on $OM$, which has no analogy in gauge theories of internal symmetries. Therefore, gravitation is geometrically "richer", having $\theta$ at its disposal; Indeed, besides the \emph{curvature $2$-forms}, due to the connections on $OM$ and $P$, we also obtain the \emph{torsion} $T:=D\theta$ on $OM$, which lacks an analogue in gauge theories of internal symmetries. 
Thanks to $\theta$ and $T$ we can construct actions {such} as \eqref{GRA} that 
do not have a direct comparison to those for gauge theories of internal symmetries {given by} \eqref{YMA} and \eqref{EMA}.

The bundle constructed over the base manifold $\mathbb{R}^4$ is trivial, i.e. $P=\mathbb{R}^4\times G$, whereas instead the base manifold $M$ of gravitational theories is in general non-contractible, thus $OM$ (or $LM$) is generally non-trivial. 
In gauge gravitational theories, the gauge group is $\text{Diff}~M$, the pure gauge group is $\left\{id_{OM}\right\}$, and the structure group is $SO(1,3)$. 
In theories of the Yang-Mills type over Minkowski space, the gauge group is $\mathscr{G}_0\ltimes \mathcal{P}$, the pure gauge group $\mathscr{G}_0$ is $\left\{S:\mathbb{R}^4\rightarrow G\right\}$, i.e., consists of local gauge transformations, and the structure group $G$ consists of global gauge transformations. Here we can see that $\mathscr{G}$ and $\mathscr{G}_0$ play opposite roles in gravitation and gauge theories of Yang-Mills type. A more detailed and general comparison of this type can be found in \cite{Trautman:1981vr}.

Lastly, we would like to comment on a common interpretation of frame fields $e_a(x)$  as gauge fields. In a gauge gravitational theory based on $(M,g,\nabla)$, we can interpret $\hat{\omega}=\sigma^*\omega$ as the gauge field and $\hat{\Omega}=\sigma^*\Omega$ as its gauge field strength, compare with \eqref{P1-1}. A frame field $e_a(x)$ is given as a local section $\sigma$ of $LM$ and its coframe field as $e^a=\sigma^*\theta^a$. The torsion is given as $\hat{T}=\sigma^*T$, where $T=D\theta$. 
Given this, one might be tempted to interpret the coframe fields as gauge fields and the torsion as the gauge field strength. However, this interpretation is problematic because $\theta$ cannot be a connection form on $LM$, as it is $\mathbb{R}^4$-valued instead of being $gl(4,\mathbb{R})$-valued as required for a connection on $LM$. 

%%%%
%%%%
%%%%
\section{TEGR as a classical gauge gravitational theory}\label{P2}
The \emph{Teleparallel Equivalent of General Relativity} (TEGR) is a gravitational theory based on the geometrical structure $\left(M,g,\nabla_{mT}\right)$, where $\nabla_{mT}$ is the \emph{metric teleparallel linear connection} defined by conditions of vanishing curvature and non-metricity $\hat{\Omega}=\nabla g=0$, which leaves only torsion to describe the non-trivial geometry of spacetime \cite{Aldrovandi:2013wha,Krssak:2018ywd}. The presence of $\nabla_{mT}$ on $(M,g)$ gives rise to a special class of orthonormal frame fields 
$\Tilde{e}_a(x)$ defined by the condition $\nabla_{mT}\Tilde{e}_a=0$, for more details see \cite[p.~425]{Fecko:2006zy}. The metric teleparallel connection 1-forms $\left(\hat{\omega}_{mT}\right)^a_b$ of $\nabla_{mT}$ vanish with respect to~$\Tilde{e}_a(x)$. An arbitrary orthonormal frame field $e_a(x)$ is obtained from $\Tilde{e}_a(x)$ by transformation $e_a(x)=\Lambda^b_a(x)\Tilde{e}_b(x)$, where $\Lambda^a_b(x)\in SO(1,3)$. Then the metric teleparallel connection 1-forms transforms accordingly 
\begin{equation}\label{telecon}
(\hat{\omega}_{mT})^a_b= \left(\Lambda^{-1}\right)^a_c(x) d\Lambda^c_b(x).
\end{equation}

The following identity between the curvature 2-forms of the Levi-Civita connection and the torsion 2-forms of the metric teleparallel connection holds \cite{Hehl:1994ue,Aldrovandi:2013wha}
\begin{equation}
-\hat{\Omega}_{ab}\wedge*\left(e^a\wedge e^b\right)=
\hat{T}^a\wedge H_a
+2 d(e^a \wedge * \hat{T}_a)
\end{equation}
where   $H_a$ is the excitation 2-form \cite{Hehl:1994ue,Krssak:2024vzo}
\begin{equation}
     H^a:=\frac{1}{2}* \left(\hat{T}^a-2e^a\wedge\iota_b\hat{T}^b+e^b\wedge\iota^a\hat{T}_b\right),
\end{equation}
and  $\iota_a$ denotes the \emph{interior product} with a frame $e_a$, i.e. $\iota_a=\iota_{e_a}$ and $\iota^a=\eta^{ab}\iota_b$.

We can then  write the TEGR action
\begin{equation}\label{TEGRA}
    \mathcal{S}_{TEGR}[e,\hat{\omega}_{mT}]:= \frac{1}{16\pi}\int \hat{T}^a\wedge H_a,
\end{equation}
which is guaranteed to be dynamically equivalent to general relativity, i.e., the field equations derived from both actions \eqref{GRA} and \eqref{TEGRA} are the same, since they differ by a total derivative term only.

TEGR as described here is a gravitational theory defined on the frame bundle $LM$, more specifically on the orthonormal bundle $OM$, as it uses ingredients, such as linear connection and torsion, available only there. Our goal here is to examine  TEGR from the fiber bundle perspective and explain whether it can be understood as a classical gauge gravitational theory. In order to do so, we need to choose the structure group of the principal $G$-bundle, determine the gauge group $\mathscr{G}$ and the pure gauge group $\mathscr{G}_0$ using the absolute elements of the theory, and ask whether this agrees with the geometrical structure of TEGR described above. We consider two choices for the structure group: the group of translations, and the Poincaré group, which includes translations as a subgroup.
%%%%
%%%%
%%%%
\subsection{Translations-only approach to TEGR}\label{P2S1}
The first choice of the structure group $G$ is the group of translations $\mathbb{T}(4)$, which is the original Cho's proposal \cite{Cho:1975dh}, and what authors of \cite{Fontanini:2018krt} and \cite{LeDelliou:2019esi}  refer to as ``translations-only" approach.
However, when it comes to this option, it was already shown in \cite{Fontanini:2018krt}  and \cite{LeDelliou:2019esi} that it faces problems. We repeat the argument here for the completeness of our exposition. 

The main issue with the translations-only approach, i.e.~with the \emph{principal $\mathbb{T}(4)$-bundle} endowed with a \emph{translational connection} $\omega_T$, is that we cannot define the torsion tensor on this bundle, as the latter exists only for frame bundles $LM$ with linear connections. Thus, the translations-only approach does not agree with the geometry of TEGR. 

A possible way out of this problem might be to work with two bundles at once, i.e.~to use the orthonormal bundle $OM$ with vanishing curvature of $\omega_{mT}$ for describing TEGR (so torsion would be defined on $OM$), and at the same time, use the principal $\mathbb{T}(4)$-bundle for developing the translations-only gauge interpretation of TEGR. 
To make this setup meaningful, one would have to link these bundles somehow. It is usually claimed, see for example \cite{Pereira:2019woq} and \cite{Aldrovandi:2013wha}, that $\omega_T$ could be identified with the canonical $1$-form $\theta$, or in other words that the translational gauge field $h^a:=\sigma_{T}^*\omega^a_T$ could be identified with an orthonormal coframe field $e^a\equiv\sigma^*\theta^a$. 

However, this identification is problematic. To see this, we first have to realize that the translational connection $\omega^a_T$ is defined on the $8$-dimensional principal $\mathbb{T}(4)$-bundle while $\theta^a$ is defined on the $10$-dimensional $OM$, thus, these bundles are not diffeomorphic. We must also realize that $e^a\equiv\sigma^*\theta^a$ is always non-vanishing for all local sections $\sigma$ of $OM$, since it is a frame field, while $h^a=\sigma_T^*\omega^a_T$ can generally vanish for the right choice of a local section $\sigma_T$ of the principal $\mathbb{T}(4)$-bundle. Therefore, these objects cannot be identified. This leads us to conclude that TEGR is not a classical gauge theory with structure group $\mathbb{T}(4)$.
%%%%
%%%%
%%%%
\subsection{Poincaré approach to TEGR \label{p2-s22}}
The second option is to consider the structure group to be the whole Poincaré group\footnote{Let us clarify that here we intend to consider the whole Poincaré group but with the connection satisfying the teleparallel constraint, i.e., having a zero curvature. This is an important difference with the Poincaré gauge theories where both the curvature and torsion are non-vanishing.
} $\mathcal{P}=SO(1,3)\ltimes \mathbb{T}(4)$, which includes both translations and proper Lorentz transformations $SO(1,3)$, and is a subgroup of  $GA(4,\mathbb{R})$. The affine group $GA(4,\mathbb{R})$ naturally acts on the affine bundle $AM$ fibers (see Appendix \ref{AppC}), making it a good principal bundle candidate for our gauge theory. 

As far as the connection is concerned, we have two possibilities on $AM$: an affine or a generalized affine connection. From Theorem~\ref{T5}, see Appendix \ref{AppC}, it follows that the use of the affine connection on $AM$ is fully equivalent to the use of the linear connection on the frame bundle $LM$.
The existence of a metric tensor with a signature $(1,3)$ reduces $GL(4,\mathbb{R})$ to the Lorentz group $SO(1,3)$,  and the vanishment of the non-metricity of the metric teleparallel connection effectively reduces a $gl(4,\mathbb{R})$-valued connection 1-form on $LM$ to just a $so(1,3)$-valued connection 1-form on $OM$. The condition $\hat{\Omega}=0$ implies that $\Omega_{mT}=0$, and  no conditions are imposed on the torsion $\hat{T}$, meaning that $T$ is generally non-vanishing. Therefore, we have obtained the known frame bundle formulation of TEGR by consistently using the whole Poincaré group. 

On the other hand, if we used the generalized affine connection, i.e.~we did not identify the $\mathbb{R}^4$-valued $1$-form $\varphi$ with $\theta$, see \eqref{A6}, then both $\varphi$ and $\theta$ would live on $OM$ independently of each other and we would have "more freedom". However, this "freedom" is not consistent with the geometrical structure of TEGR because no such $\varphi$ exists there (unless $\varphi=\theta$, the affine connection case). Thus, a generalized affine connection cannot be used. 

After establishing the viability of the Lorentz group as the structure group and the use of the linear connection, let us now identify the absolute elements of the theory. Here we should recall that the TEGR field equations are just Einstein field equations rewritten in terms of the metric teleparallel geometric quantities, and hence they determine only the metric tensor, or equivalently, they determine the orthonormal frame field $e$ only up to a local Lorentz transformation \cite{Aldrovandi:2013wha,Krssak:2018ywd}. 
The variation of the action with respect to the metric teleparallel connection $\hat{\omega}_{mT}$ vanishes identically \cite{Krssak:2015lba,Golovnev:2017dox}, and hence any metric teleparallel connection automatically satisfies the corresponding field equations for the connection. 

This raises the question of whether a metric teleparallel connection should be regarded as a dynamical variable within Trautman’s framework. Due to the triviality of the connection field equations in TEGR, it is common to view the metric teleparallel connection as non-dynamical \cite{Aldrovandi:2013wha,Krssak:2018ywd}. It is important to emphasize, however, that non-dynamical is usually understood here in the sense of not influencing the physical dynamics of the theory, i.e., not affecting its solutions. In contrast, within Trautman’s framework, absolute elements are non-dynamical geometric objects in a stronger sense: they do not possess field equations at all, such as the Minkowski metric $\eta$ in Yang–Mills theory or the canonical 1-form $\theta$ in gravitational gauge theories. It is therefore worthwhile to examine both cases, in which the metric teleparallel connection is treated as dynamical or as non-dynamical\footnote{We would like to emphasize that we are asking here whether $\omega_{mT}$, obtained through \eqref{LM8} from $\hat{\omega}_{mT}$, is a dynamical variable.}.

We start with the first option, when a metric teleparallel connection is treated as a dynamical object. 
Then TEGR can be understood as a classical gauge theory within Trautman's framework. The only absolute element of the theory is the canonical 1-form $\theta$, which yields Diff~$M$ as the gauge group, and the pure gauge group is trivial, see Section~\ref{P1}.

The second option is to treat the metric teleparallel connection as a non-dynamical variable of the theory. 
This immediately represents a substantial departure from Trautman's definition of a classical gauge theory by relaxing the requirement of a dynamical connection, and leads to the question whether the connection should be treated as an absolute element. We are then asking whether it is possible to view TEGR as a theory with two absolute elements: the canonical 1-form $\theta$, and the connection 1-form $\omega_{mT}$. Then the gauge transformations are given as such automorphisms $u\in$ Aut~$OM$ that preserve both absolute elements, i.e., they are given by conditions
\begin{equation}\label{P2-1}
    u^*\theta=\theta, \quad\quad u^*\omega_{mT}=\omega_{mT}.
\end{equation}
The first condition implies that the gauge group is isomorphic to $\text{Diff}~M$ and the pure gauge group is trivial, see the paragraph above equation \eqref{GRA}. The second condition further restricts $\text{Diff}~M$ to a subgroup $\mathscr{G}$. As discussed in Appendix~\ref{appprob}, this leads to the problem that the gauge group $\mathscr{G}$, cannot be determined, and is possibly not unique.

This insight motivates us to further modify Trautman’s framework by requiring that the metric teleparallel connection, even if it is not a dynamical variable, is not treated as an absolute element. Then again, TEGR can be viewed as a classical gauge theory with a gauge group Diff~$M$, since the only absolute element is the canonical 1-form $\theta$. 
%%%%
%%%%
%%%%

\subsection{Poincar\'{e} approach from the TEGR viewpoint \label{subsec-bridge}}
Let us now explain how our approach is related to the presentation of TEGR as a gauge theory in  \cite{Aldrovandi:2013wha,Krssak:2018ywd,Pereira:2019woq}.  As already mentioned in Section~\ref{P2S1}, the problem of the ``translation-only" approach lies in identifying the translational gauge field with the coframe. In the language of   \cite{Aldrovandi:2013wha,Krssak:2018ywd,Pereira:2019woq}, this corresponds to a translational covariant derivative $h_\mu$ acting on matter fields $\Psi$ as 
\begin{equation}
 h_\mu \Psi=h^a{}_\mu \partial_a \Psi,
\end{equation}
where $\partial_a$ are partial derivatives with respect to the tangent space coordinates \footnote{To be more precise, these coordinates are not defined on $TM$ but on the principal $T(4)$-bundle, but we keep the terminology of references \cite{Aldrovandi:2013wha,Krssak:2018ywd,Pereira:2019woq}.} $\chi^a$. The components of $h_\mu$ are then written as
\begin{equation}\label{covtett1}
	h^a{}_\mu =\partial_\mu \chi^a  
	+B^a{}_\mu,
\end{equation}
which is then identified with the components of the frame field $e^a{}_\mu$. However, since the frame fields are non-degenerate, then they cannot be identified with the components of the connection, which may vanish for some choice of the local section. A straightforward example when this happens is the case when the translational field strength is zero, and then the connection is pure-gauge and hence guaranteed to vanish for some choice of the local section of the principal $\mathbb{T}(4)$-bundle. 

Note that the problem as described above corresponds to the original translation-only approach to teleparallel gravity--known as the non-covariant or pure-tetrad formulation \cite{Krssak:2018ywd}--where the only variable is the frame field. In the covariant formulation, the metric teleparallel connection $(\hat{\omega}_{mT})^a{}_b$ appears as an additional variable to the frame field ensuring appropriate tensorial behavior of the torsion tensor \cite{Krssak:2018ywd,Krssak:2024xeh}. However, in the  gauge treatment of this covariant teleparallel gravity  \cite{Aldrovandi:2013wha,Krssak:2018ywd,Pereira:2019woq}, 
the teleparallel connection $(\hat{\omega}_{mT})^a{}_b$  is added to the translational connection \eqref{covtett1}  as 
\begin{equation}\label{covtett}
h^a{}_\mu =\partial_\mu \chi^a  
+B^a{}_\mu+(\hat{\omega}_{mT})^a{}_{b\mu} \chi^b,
\end{equation}
which is then identified with a general orthonormal frame. 

While the problem of identification of the connection with the frame from the non-covariant approach remain, we face here an additional difficulty of keeping the interpretation of the right-hand-side of \eqref{covtett} as a translational covariant derivative since it now explicitly contains the teleparallel spin connection, which is related to the Lorentz transformations. The teleparallel spin connection $\hat{\omega}_{mT}$ is added to \eqref{covtett}  after ``gauging" the translations.   The motivation for this is to have gravity described in terms of torsion only, but to be able to describe it in terms of arbitrary {orthonormal frames} in a covariant way.

Our suggestion, on the other hand, is to use the affine connection $\Tilde{\omega}=\Tilde{\omega}^a_bE^b_a+\Tilde{\omega}^aE_a$ on the affine bundle $AM$\footnote{This affine approach is also discussed in \cite[Section 14.1.2]{CANTATA:2021asi} but there the translational gauge approach is uphold. A similar situation was analyzed in \cite{Grignani:1991nj} and \cite{Koivisto:2019ejt}, where the conclusions supported the Lorentz gauge theory interpretation.}, since the relation \eqref{covtett} corresponds to the covariant derivative of $\chi^a$ with respect to $\sigma_{AM}^*\Tilde{\omega}$. It turns out \cite{KN}, see Appendix \ref{AppC}, that the pullback of the affine connection $\Tilde{\omega}$  from $AM$ to $LM$ by the bundle homomorphism $\gamma$,  then decomposes as\footnote{However, $\gamma^*\Tilde{\omega}$ is not a connection on $LM$, see \eqref{A6}, since it is a $ga(4,\mathbb{R})$-valued 1-form and not a $gl(4,\mathbb{R})$-valued 1-form as the connection on $LM$ is required to be by definition.} $\gamma^*\Tilde{\omega}=\omega+\theta$, where $\omega$ is the linear connection on $LM$ and $\theta$ is the canonical 1-form. This decomposition allows us to see that the translational part $\Tilde{\omega}^a$ of the affine connection $\Tilde{\omega}$, which is related to the canonical 1-form $\theta$ as $\theta^a=\gamma^*\Tilde{\omega}^a$, does not give rise to a translational gauge field on $M$ since $\theta$ is not a connection 1-form on $LM$--this was already pointed out by Sardanashvily \cite{Sardanashvily:1983xbn, Sardanashvily:1984dx, Ivanenko:1987ka} in the case of Poincaré gauge theories. Moreover, as shown by Dass  \cite{Dass:1984qk}, this observation indeed allows us to obtain coframe fields $e^a$ through $\Tilde{\omega}^a$ as $e^a=\sigma_{LM}^*\left(\gamma^*\Tilde{\omega}^a\right)$, but it does not mean that frame fields are translational gauge fields.

To conclude, the analysis provided above implies that the left-hand side of \eqref{covtett} is not a frame field since the right-hand side is the covariant derivative of $\chi^a$ with respect to $\sigma_{AM}^*\Tilde{\omega}$ and that $B^a{}_\mu$ is not a translational gauge field since affine connections $\Tilde{\omega}$ (or equivalently linear connections $\omega$) do not give rise to translational gauge fields on $M$. Thus, TEGR cannot be consistently viewed as a classical gauge theory with a translational connection, but rather as a classical gauge theory with affine or linear connections. 
%%%%
%%%%
%%%%
\section{Discussion and conclusions} 
This work examined the gauge structure of the Teleparallel Equivalent of General Relativity (TEGR) within the framework of principal bundles. The primary objective was to assess whether TEGR can be interpreted as a classical gauge theory in the sense introduced by Trautman, where gauge theories are characterized by a connection on a principal bundle with the structure group $G$, and a set of absolute elements--geometric objects without their own field equations in the considered gauge theory.

The advantage of Trautman's framework is that it allows for treating a wide range of situations, including theories of Yang-Mills type with both trivial and non-trivial principal bundles, as well as gravitational theories. Moreover, it makes a clear distinction between the structure groups and gauge groups, which turns out to be better and more transparent than the usual discussion of global and local gauge symmetries found commonly in the physics literature. We have {reviewed how this framework is} applied to Yang–Mills theories, as well as to general relativity and other classical gauge theories of gravity, which can be viewed as a gauge theory with the structure group $G=SO(1,3)$--or more generally $G=GL(4,\mathbb{R})$--and the canonical 1-form on the frame bundle playing the role of the absolute element, resulting in the group of spacetime diffeomorphisms being its gauge symmetry.

We have then explored how to accommodate TEGR within the framework of Trautman, where we have started with a discussion of the choice of the structure group. We have first addressed the problems of formulating TEGR theory as a classical gauge theory with the structure group being the group of translations, where we have agreed with the previous results \cite{Fontanini:2018krt,LeDelliou:2019esi,Huguet:2021roy}, as well as \cite{Dass:1984qk,Sardanashvily:1983xbn}, arguing that the identification of the translational connection with the orthonormal coframes is problematic due to their different geometric nature.

Instead, we have argued that TEGR can be formulated using the affine bundle $AM$ with the Poincaré group as the structure group, which is fully equivalent to using the linear connection on the orthonormal frame bundle $OM$ with the structure group $G = SO(1,3)$ and the canonical 1-form $\theta$ due to the existence of bundle homomorphism discussed in Appendix~\ref{AppC}\footnote{Note that an alternative approach has been proposed in \cite{LeDelliou:2019esi,Huguet:2020ler,Huguet:2021roy}, where the connection and canonical form are treated as components of a Cartan connection \cite{Wise:2006sm,Westman:2014yca,Catren:2014vza}.}. The orthonormal frame bundle $OM$ is the natural geometric framework within which both TEGR and general relativity are usually formulated.

However, we have encountered a crucial difference that sets TEGR apart from other gravitational gauge theories: the triviality of the field equations for connection\footnote{The same situation is also happening in the so-called symmetric and general teleparallel equivalents of general relativity \cite{Nester:1998mp, BeltranJimenez:2019odq}.}. This then led us to question whether the metric teleparallel connection should be considered to be a dynamical or non-dynamical variable.

When we treat the connection as a dynamical variable--which is possible since any metric teleparallel connection trivially satisfies the null field equations--the only absolute element is the canonical $1$-form. The gauge group is $\mathscr{G} \approx \text{Diff}~M$, while the pure gauge group $\mathscr{G}_0$ is trivial, and TEGR can be understood as a classical gauge theory of gravity within Trautman's framework.

On the other hand, when the connection is treated as a non-dynamical variable,  it becomes necessary to modify Trautman’s definition of a classical gauge theory by relaxing the requirement that the connection be dynamical, which raises the question of whether the connection should be regarded as an absolute element. We have demonstrated that if the metric teleparallel connection is considered to be the absolute element along with the canonical 1-form $\theta$,  we find that the gauge group of TEGR becomes some subgroup of diffeomorphisms. This led to a potential problem: since the connection is not given uniquely, the gauge group cannot be uniquely determined, raising the possibility that it may depend on the physical situation under consideration. Therefore, in order to accommodate TEGR, if the connection is considered non-dynamical, it is necessary not to treat it as an absolute element, which represents a further modification of Trautman’s framework.

In both cases, the only absolute element is the canonical 1-form $\theta$, and TEGR can be viewed as a classical gauge theory on the orthonormal frame bundle $OM$ with the structure group $G = SO(1,3)$, and the gauge group given by spacetime diffeomorphisms, $\mathscr{G} \approx \text{Diff}~M$, while the pure gauge group $\mathscr{G}_0$ is trivial. The gauge structure of TEGR then agrees with that of general relativity.

Let us now conclude with some comments on the role of translations in TEGR--and more generally in gravitational theories\footnote{We consider here only gravitational theories with vanishing non-metricity, which are naturally formulated on $OM$, since we would like to compare them with TEGR. Inclusion of non-metricity is rather straightforward by generalizing the geometrical setting from $OM$ to $LM$.}--which highlights a crucial distinction between the notion of gauge symmetry in the framework of principal bundles and the “physicist’s” viewpoint where (local) gauge symmetries are seen as ``localized"  global symmetries of the action. While in Yang–Mills-type theories these two perspectives are effectively equivalent, difficulties arise when attempting to apply the latter to translations and gravity.

This issue becomes evident when one promotes translations of the form $x^\mu \rightarrow x^\mu +\varepsilon\xi^\mu$ to so-called local translations $x^\mu \rightarrow x^\mu + \varepsilon\xi^\mu(x)$, where $\varepsilon$ is infinitesimally small. However, the latter are not translations in the proper sense, but rather diffeomorphisms. In this framework, then, what is commonly referred to as the “localization” of translations is in fact a replacement of translations by diffeomorphisms. While this distinction may often be benign, it becomes problematic--and often a source of confusion--when attempting to incorporate translations into the language of principal bundles.

From the principal bundle perspective, the role of translations is more subtle. Gravitational theories of interest can be formulated using the affine bundle $AM$, with the Poincaré group as the structure group, which includes translations as a subgroup. However, due to the existence of bundle homomorphisms between the affine bundle $AM$ and the orthonormal frame bundle $OM$, this formulation is equivalent to using the linear connection on $OM$ with the structure group $SO(1,3)$ and the canonical 1-form $\theta$; see Appendix~\ref{AppC}. Although $\theta$ is indirectly related to translations through the bundle homomorphisms, it does exist independently on the orthonormal frame bundle $OM$, where no translation symmetry is present, and where it gives rise to torsion. This explains why, in TEGR and other gravitational theories formulated using the formalism of orthonormal frames, we are usually concerned with local Lorentz transformations of coframes only.
 
%%%%
%%%%
%%%%
\section*{Acknowledgements} 
We are grateful to Igor Khavkine for the fruitful discussions during the 45th Winter School Geometry and Physics 2025 in Srní, Czech Republic, and to Eric Huguet and Morgan Le Delliou. The work of M.K.~was funded through  SASPRO2 project \textit{AGE of Gravity: Alternative Geometries of Gravity}, which has received funding from the European Union's Horizon 2020 research and innovation programme under the Marie Skłodowska-Curie grant agreement No. 945478, and VEGA grant 	1/0565/25. E.B.~is supported by the GA\v{C}R grant PIF-OUT 24-10634O.
%%%%
%%%%
%%%%
\appendix
\setcounter{equation}{0}\renewcommand\theequation{A\arabic{equation}}
\section{Principal G-bundles and connections}\label{AppA}
{A \emph{fiber bundle} is the structure  denoted as $\pi: \mathcal{B}\rightarrow M$, consisting of the quadruple $(\mathcal{B}, M, \pi, F)$, where $\mathcal{B}$ is a smooth $(m+n)$-dimensional manifold called the \emph{total space}, assigned to a smooth $n$-dimensional manifold $M$, known as the \emph{base}, and $\pi$ is a surjective map between $\mathcal{B}$ and $M$ referred to as the \emph{canonical projection}. The manifold $F$ is a smooth $m$-dimensional manifold called the \emph{typical fiber}, such that for each point $x\in M$ the preimage\footnote{Since $\pi$ is just a surjective map, $\pi^{-1}(x)$ denotes preimages of $x$ and not the inverse map of $\pi$.} $F_x=\pi^{-1}(x)$ called the \emph{fiber over} $x$ is a smooth submanifold of $\mathcal{B}$  diffeomorphic to $F$. 

Moreover, the fiber bundle has a local product structure, meaning that there exists covering $\mathcal{O}_\alpha$ of the base $M$ such that on $\pi^{-1}(\mathcal{O}_\alpha)$  the fiber bundle can be written as a direct product $\pi^{-1}(\mathcal{O}_\alpha)\rightarrow \mathcal{O}_\alpha\times F$.  The product bundles are those that can be written as $\mathcal{B}=M\times F$, and fiber bundles equivalent to product bundles are called \emph{trivial}.}
\begin{theorem}\label{T1}
If the base $M$ of $\pi: \mathcal{B}\rightarrow M$ is contractible, then $\mathcal{B}$ is trivial.
\end{theorem}

A \emph{local section} of the fiber bundle $\pi: \mathcal{B}\rightarrow M$ is a smooth map $\sigma: \mathcal{O}\rightarrow \mathcal{B}$, where $\mathcal{O}\subset M$, such that $\pi\circ \sigma=id_{\mathcal{O}}$. If $\mathcal{O}=M$, then the section is called \emph{global}. {While local sections always exist, existence of the global sections is not guaranteed.}

{A \emph{principal $G$-bundle} is a fiber bundle $(P,M,\pi,G)$  on which we have additionally defined the right vertical action of a Lie group $G$, which satisfies the following properties:}
\begin{equation}
    R_g: P\rightarrow P \quad R_{gh}=R_h\circ R_g \quad R_e=id \quad \pi\circ R_g=\pi,
\end{equation}
where $F$ is required to be diffeomorphic to $G$ as a manifold. {The group action is required to be} \emph{free} (all stabilizers are trivial) and \emph{transitive} (any two points in a single fiber can be joined by the action) within a fiber. 

A \emph{(principal Ehresmann) connection} is an additional structure defined on a principal $G$-bundle{,  described by} a $\mathcal{G}$-valued \emph{connection $1$-form} $\omega=\omega^i E_i$, where $E_i$ is a basis in the Lie algebra $\mathcal{G}$ of $G$, and $\omega^i$ are ordinary 1-forms on $P$. 
Using the connection, one can introduce the \emph{exterior covariant derivative} $D$ of a $p$-form $\alpha$ on $P$, 
which defines the \emph{curvature $2$-form} $\Omega=\Omega^iE_i$ of the connection $1$-form $\omega$ 
as
\begin{equation}\label{E4}
      \Omega:=D\omega.
 \end{equation}
A \emph{function $\Phi$ of type $\rho$} is an equivariant function on $P$ defined as
 \begin{equation}\label{E5}
     \Phi: P\rightarrow (V,\rho) \quad\quad \Phi\circ R_g=\rho(g^{-1})\circ\Phi,
 \end{equation}
where $(V,\rho)$ is a linear space, in which a representation $\rho$ of the Lie group $G$ is available. 
%%%%
%%%%
%%%%
\section{Frame bundle and a linear connection}\label{AppB}
\setcounter{equation}{0}\renewcommand\theequation{B\arabic{equation}}
The \emph{frame bundle} $\pi: LM\rightarrow M$ is a \emph{principal $GL(n,\mathbb{R})$-bundle} {with fiber $F_x$ consisting} of all frames spanning $T_xM$. Such a relation is called \emph{soldering}, and since $T_xM$ exists at $M$ without any additional structures, the frame bundle is a special case of a principle bundle canonically associated with $M$.
Local sections $\sigma$ of $LM$ over $\mathcal{O}\subset M$ give rise to frame fields $e_a(x)$ on $\mathcal{O}$.
Due to the soldering property of $LM$, there  exists the so-called \emph{soldering} or \emph{canonical $1$-form} $\theta=\theta^aE_a$, where $E_a$ is a basis in $\mathbb{R}^n$, defined at the point $e\in LM$ as
\begin{equation}\label{LM1}
    \langle \theta^a_e,w \rangle:=\langle e^a,\pi_*w \rangle,
\end{equation}
where $e^a$ is the dual basis (coframe) to $e_a$, for $a=1,...,n=\text{dim}~M$, and $w$ is a tangent vector at the point $e$. The canonical $1$-form is related to a coframe field in the region $\mathcal{O}$ by the following relation: 
\begin{equation}\label{LM3}
    e^a=\sigma^*\theta^a.
\end{equation}

As an additional structure, we introduce a connection $\omega$ on $LM$, called a \emph{linear connection}. Besides the curvature 2-form $\Omega = D\omega$, which is defined on any principal $G$-bundle with a connection, the linear connection also gives rise to the  \emph{torsion $2$-form} $T$, defined uniquely on $LM$ due to the presence of both the canonical 1-form $\theta$ and the linear connection $\omega$,  as
\begin{equation}\label{LM4}
  T:=D\theta.  
\end{equation}

The forms $\omega$, $\Omega$ and $T$, living on the total space $LM$, are related to objects in the theory of linear connections $(M, \nabla)$ on the base $M$: the \emph{linear connection $1$-form} $\hat{\omega}$, the \emph{curvature $2$-form} $\hat{\Omega}$, and the \emph{torsion $2$-form} $\hat{T}$. These are related via the pullback by a local section $\sigma$ of $LM$, given as
  \begin{equation}\label{LM6}
    \hat{\omega}=\sigma^*\omega,  \quad\quad \hat{\Omega}=\sigma^*\Omega, \quad \quad \hat{T}=\sigma^*T.
 \end{equation}
On the other hand, given the linear connection 1-form $\hat{\omega}$ of $\nabla$ with respect to a frame field $e_a(x)$ in a region $\mathcal{O}\subset M$, 
the corresponding linear connection 1-form $\omega$ on $\pi^{-1}(\mathcal{O})\subset LM$  is obtained as
\begin{equation}\label{LM7}
\omega=y^{-1}\left(\pi^*\hat{\omega}\right)y+y^{-1}dy,
\end{equation}
or, in the index notation, expressed as
\begin{equation}\label{LM8}
  \omega^a_b=\left(y^{-1}\right)^a_c\left(\pi^*\hat{\omega}^c_d\right)y^d_b+\left(y^{-1}\right)^a_cdy^c_d,
\end{equation}
where $\left(x^i,y^a_b\right)$ are local coordinates in $\pi^{-1}(\mathcal{O})$. The coordinates $y^a_b$ describe an arbitrary frame $E_a$ in the fiber $\pi^{-1}(x)$ 
obtained from the reference frame $e_a\in\pi^{-1}(x)$ as $E_a=y^b_ae_b$.
%%%%
%%%%
%%%%
\section{Affine bundle and an affine connection}\label{AppC}
\setcounter{equation}{0}\renewcommand\theequation{C\arabic{equation}}
An \emph{affine group} $GA(n,\mathbb{R})$ is a semidirect product of $GL(n,\mathbb{R})$ and the group of translations  ${\mathbb{T}(n):=\left(\mathbb{R}^n,+\right)}$, i.e.,  $GA(n,\mathbb{R})=GL(n,\mathbb{R})\ltimes \mathbb{T}(n)$.
There exists an injective homomorphism $\Sigma:GA(n,\mathbb{R}) \rightarrow GL(n+1,\mathbb{R})$ allowing us to represent $(A,v)\in GA(n,\mathbb{R})$, where $A\in GL(n,\mathbb{R})$ and $v\in \mathbb{T}(n)$, by a matrix in $GL(n+1,\mathbb{R})$ as
\begin{equation}\label{A1}
    (A,v)\longleftrightarrow
\begin{pmatrix} 
A & v  \\
 0 & 1 \\
\end{pmatrix}.
\end{equation}
The Lie algebra of $GA(n,\mathbb{R})$ is given by $ga(n,\mathbb{R})=gl(n,\mathbb{R})+\mathbb{R}^n$, where $+$ denotes the semidirect sum. An element $(X,\chi)\in ga(n,\mathbb{R})$ can be represented by a matrix from $gl(n+1,\mathbb{R})$ as
\begin{equation}\label{A2}
    (X,\chi)\longleftrightarrow
\begin{pmatrix} 
X & \chi  \\
 0 & 0 \\
\end{pmatrix}.
\end{equation}
There exist a homomorphism $\alpha$ of $\mathbb{T}(n)$ into $GA(n,\mathbb{R})$, a homomorphism $\beta$ of $GA(n,\mathbb{R})$ onto $GL(n,\mathbb{R})$ and a homomorphism $\gamma$ between $GL(n,\mathbb{R})$ and $GA(n,\mathbb{R})$, given as
\begin{equation}\label{A3}
    v ~ \overset{\alpha}{\longrightarrow} ~ \begin{pmatrix} 
\mathds{1} & v  \\
 0 & 1 \\
\end{pmatrix}, \quad\quad
\begin{pmatrix} 
A & v  \\
 0 & 1 \\
 \end{pmatrix}
 ~ \overset{\beta}{\longrightarrow}~ A, \quad\quad
 A ~ \overset{\gamma}{\longrightarrow} ~\begin{pmatrix} 
A & 0  \\
 0 & 1 \\
\end{pmatrix},
\end{equation}
and hence $\beta\circ \gamma$ is the identity automorphism of $GL(n,\mathbb{R})$, and $(A,v)\leftrightarrow \alpha(v)\gamma(A)$.

An \emph{affine bundle} $\pi: AM\rightarrow M$ is a \emph{principal $GA(n,\mathbb{R})$-bundle} with fiber $F_x$ consisting of all affine frames spanning the affine tangent space $A_xM$.
The group homomorphisms $\beta$ and $\gamma$ between $GA(n,\mathbb{R})$ and $GL(n,\mathbb{R})$ give rise to bundle homomorphisms between $AM$ and $LM$.

A connection defined on $AM$ is called a \emph{generalized affine connection}. The corresponding connection 1-form $\Tilde{\omega}$, which takes values in $ga(n,\mathbb{R})$, can therefore be decomposed as 
\begin{equation}\label{A4}
\Tilde{\omega}=\Tilde{\omega}^a_bE^b_a+\Tilde{\omega}^aE_a,
\end{equation}
where the first term is $gl(n,\mathbb{R})$-valued and the second $\mathbb{R}^n$-valued. The curvature $2$-form $\Tilde{\Omega}$ of $\Tilde{\omega}$ can be decomposed as
\begin{equation}\label{A5}
\Tilde{\Omega}=\Tilde{\Omega}^a_bE^b_a+\Tilde{\Omega}^aE_a.
\end{equation}
It can be shown \cite[p.~127]{KN} that a pullback of $\Tilde{\omega}$ by $\gamma: LM\rightarrow AM$ yields a $ga(n,\mathbb{R})$-valued 1-form on $LM$, which is not a connection 1-form on $LM$, but is given by
\begin{equation}\label{A6}
    \gamma^*\Tilde{\omega}=\omega+\varphi,
\end{equation}
where $\omega$ is a $gl(n,\mathbb{R})$-valued 1-form defining the linear connection on $LM$, and $\varphi$ is just an $\mathbb{R}^n$-valued 1-form on $LM$, which does not define a connection on $LM$. Similarly,  for the pullback of the curvature $\Tilde{\Omega}$ by $\gamma$ we get a decomposition 
\begin{equation}\label{A7}
    \gamma^*\Tilde{\Omega}=\Omega+D\varphi,
\end{equation}
where $\Omega$ is the curvature of the linear connection $\omega$ on $LM$, and $D$ is the exterior covariant derivative with respect to the linear connection $\omega$.

A generalized affine connection is called an \emph{affine connection} iff $\varphi$ is chosen to be the canonical 1-form $\theta$. The affine and linear connections are closely related through the following theorem \cite[p.~129]{KN}:
\begin{theorem}\label{T5}
The homomorphism $\beta: AM\rightarrow LM$ maps every affine connection into a linear connection, and moreover, gives a 1-1 correspondence between the set of affine connections and the linear connections.
\end{theorem}
For an affine connection, equations \eqref{A6} and \eqref{A7} take the form
\begin{equation}\label{A8}
   \gamma^*\Tilde{\omega}=\omega+\theta, \quad\quad \gamma^*\Tilde{\Omega}=\Omega+T.
\end{equation}
%%%%
%%%%
%%%%
\section{Non-dynamical teleparallel connection \label{appprob}}
\setcounter{equation}{0}\renewcommand\theequation{D\arabic{equation}}
Let us expand on problems with non-dynamical connection discussed at the end of Section~\ref{p2-s22},  where we follow \cite{Trautman:1979fg}. We explore which gauge group arises when we impose condition \eqref{P2-1}.
While the first condition is standard--the canonical 1-form $\theta$ is an absolute element--the second condition, invariance of a metric teleparallel connection, is usually an additional requirement imposed only when one is interested in symmetries of the connection.

Infinitesimally, we are asking that the Lie derivative along $Z$, a tangent vector to the one parameter group $u_t$, $t\in \mathbb{R}$, is zero:
\begin{equation}
\mathcal{L}_Z \theta = 0, \quad \mathcal{L}_Z \omega_{mT} = 0 .
\end{equation}
A few consequences are immediate, using Cartan's formula $\mathcal{L}_Z = [d,\iota_Z] = d \iota_Z + \iota_Z d$: 
\begin{align}
\mathcal{L}_Z \theta 
=\iota_Z T_{mT} + (d + \omega_{mT}\wedge ) \iota_Z \theta - (\iota_Z \omega_{mT}) \wedge \theta , \label{theta-inv}
\end{align}
and 
\begin{align}
\mathcal{L}_Z \omega_{mT} =
(d + \omega_{mT}\wedge ) \iota_Z \omega_{mT}  {{-\left(\iota_Z \omega_{mT}\right)\wedge\omega_{mT} }},\label{omega-inv}
\end{align}
since the curvature of the metric teleparallel connection is zero. For $Z$ we shall now use that, since
\begin{equation}
0 \rightarrow \text{Aut}_0~OM \xrightarrow{i} \text{Aut}~OM \xrightarrow{j} \text{Diff}~M \rightarrow 0
\end{equation} 
is exact,\footnote{There is an analogous short exact sequence for the gauge group and the pure gauge group:
\[
0 \rightarrow \mathscr{G}_0 \rightarrow \mathscr{G} \rightarrow \mathscr{G}/\mathscr{G}_0 \rightarrow 0 .
\]
} with $i$ the canonical inclusion and $j$ a projector, and there is a splitting of the sequence $L : \text{Diff}~M  \rightarrow \text{Aut}~OM $, then $Z$ is rather given by the generator of $v_t$ which is in the image, $v_t = j(u_t)$.
In other words, if we have a frame $e \in OM$ and $\xi_{v}$ and $\xi_{u}\equiv Z$ are generating vector fields for $v_t$ and $u_t$, then the following holds: 
\begin{equation}
\langle Z, \theta\rangle = \langle \pi_* \xi_u, e \rangle = \langle  \xi_v, e \rangle.
\end{equation}
The sole existence of the splitting $L$ and that $L(v_t)=u_t$ indicates that the pure gauge group $\mathscr{G}_0$ is the identity and the latter combines with $\text{Diff}~M$ to form the gauge group $\mathcal{G}$.   
What would the invariance condition \eqref{theta-inv} imply in the case of general relativity then? 

Let us express $Z$ in the dual basis to $\theta$ and $\omega_{mT}$: 
\begin{equation}
\iota_Z \theta^a = X^a, \quad \iota_Z \left(\omega_{mT}\right)^{a}{}_b = Y^{a}{}_b.
\end{equation} 
From now on, we suppress these indices to avoid the heaviness of notation when the meaning is clear. In general relativity, where the question is about finding  
$Z$ for which $\mathcal{L}_Z \theta = 0$ (and the torsion is zero), the condition, obtained by composing with a local section, is just 
\begin{equation}
D_{LC} X = Y \wedge \theta .
\label{GRcond}
\end{equation}
Hence, the component $Y$ depends only on the horizontal part ($X$). This is in agreement with the fact that $\mathscr{G}_0$ is trivial.

Coming back to the investigation for TEGR, under the request that $\omega_{mT}$ is invariant, composing \eqref{theta-inv} and \eqref{omega-inv} with a local section from the right, we get:
\begin{align}
\iota_Z T_{mT} + D_{mT} X = Y \wedge \theta, \quad D_{mT} Y = 0
\label{mTEGRcond}
\end{align}
In order to solve these equations, we notice that a first issue arises from the fact that the metric teleparallel connection is not determined by the field equations. Therefore the above differential equations contain too many undetermined variables. So let us assume that $\omega_{mT}$ is given and analyse the solutions to the above pair of equations. The equation on the left is the counterpart to \eqref{GRcond} for non-zero torsion, and can be solved to express $Y$ in terms of the unknown $X$, which still implies that the pure gauge group is trivial. Then the equation on the right imposes further constraints on $Y$ as a function of $X$. Hence, the gauge group is strictly smaller than the group of diffeomorphisms, because $\mathscr{G}/\mathscr{G}_0 \subset \text{Diff}~M$. So, compared to GR, the gauge group is smaller in a principal bundle formulation of TEGR, with the additional request that the connection is invariant.

In the analysis provided above, we have seen that a modification of Trautman's framework--relaxing the original assumption of a dynamical connection--leads to complications when one wants to determine the gauge group of TEGR. These complications are caused by the dependence of the gauge group on an unknown metric teleparallel connection and can even lead to non-uniqueness or triviality of the gauge group. The source of all this is the treatment of the metric teleparallel connection as an absolute element.

\,
%%%%
%%%%
%%%%
\bibliography{references}\addcontentsline{toc}{section}{References}
\bibliographystyle{Style}

\end{document}